\documentclass[aps,prd,twocolumn,groupedaddress,amssymb,eqsecnum,showpacs]{revtex4}
\usepackage{graphicx}
\usepackage{bm}

\begin{document}

\preprint{CMU-HEP-02-02}

\title{RS1 Cosmology as Brane Dynamics in an AdS/Schwarzschild Bulk}

\author{Hael Collins}
\email{hael@cmuhep2.phys.cmu.edu}
\author{R.~Holman}
\email{holman@cmuhep2.phys.cmu.edu}
\author{Matthew R.~Martin}
\email{mmartin@cmu.edu}
\affiliation{Physics
Department\\Carnegie Mellon University\\Pittsburgh PA 15213}

\date{\today}

\begin{abstract}
We explore various facets of the cosmology of the Randall-Sundrum scenario
with two branes by considering the dynamics of the branes moving in a bulk
AdS/Schwarzschild geometry.  This approach allows us both to understand in
more detail and from a different perspective the role of the stabilization of
the hierarchy in the brane cosmology, as well as to extend to the situation
where the metric contains a horizon.  In particular, we explicitly determine
how the Goldberger-Wise stabilization mechanism perturbs the background bulk
geometry to produce a realistic cosmology.
\end{abstract}


\maketitle

\section{Introduction}\label{sec:Intro}

The cosmology of Randall-Sundrum (RS) \cite{rsa,rsb} models is of great
interest for a variety of reasons, not least of which is that these models
allow us to understand some long-standing problems of particle physics, most
notably the hierarchy problem of mass scales. 

Kraus \cite{kraus}, (see also \cite{cosmologies,chambreall,others}) has made
the interesting observation that for the so-called RS2 models \cite{rsb},
brane cosmology can be found by solving the Israel junction conditions
\cite{israel} for a domain wall moving in an AdS/Schwarzschild (AdSS) bulk
geometry, which, by Birkhoff's theorem in five dimensions, is the unique
spherically symmetric solution to the bulk Einstein equations with a bulk
cosmological constant $\Lambda$.

While the RS2 alternative to compactification is extremely interesting from
many points of view, such as the AdS/CFT correspondence \cite{adscft}, it
could be argued that from a particle physics perspective, RS1 geometries
\cite{rsa} are more useful, allowing as they do for solutions to the hierarchy
problem.  The cosmology of RS1 geometries has been studied extensively in the
literature \cite{langlois,csaki1,csaki2,cline,ira}, and a number of
interesting results have been found.  The main difference between RS1 and RS2
cosmologies is that the existence of {\it two\/} branes (the TeV and Planck
branes) in the RS1 case allows the inter-brane separation to become a
dynamical degree of freedom, the so-called {\it radion\/}.  Depending on
whether or not the radion has been stabilized in some way, such as with the
Goldberger-Wise mechanism (GW) \cite{gw} or in some other way, the
cosmological consequences and even consistency of RS1 models can be quite
different.

Our purpose in this work is to treat the cosmology of RS1 models using the
Kraus approach.  That is to say, we consider the two branes to be moving in an
AdSS bulk geometry  and ask once again what is happening to the intrinsic
brane geometry as specified through the Israel conditions.  There are (at
least!) two reasons for doing this.  One is that this will allow us to
understand some of the previously known results from a different perspective.
In particular, in this approach, there is {\it no\/} radion mode in the bulk
metric!  The radion appears as the geodesic relative distance between the
branes.  This naturally leads us to ask how the equation of motion for the
radion gets generated, and in particular, how the GW mechanism gets
implemented.  There should also be some differences between the so-called
Gaussian normal coordinates used in most of the works on RS1 cosmology and
this method, since in the latter, the bulk is {\it static\/}, at least to a
first approximation; all the cosmological evolution occurs on the branes. 
This approach also allows the role of the GW mechanism in producing an
acceptable cosmology to be seen in more detail than for the effective action
used in the Gaussian normal formulation.  Finally, we can make contact with
the ideas of holography \cite{holo} by placing a horizon in the bulk geometry
and seeing how that influences the cosmology and stabilization mechanisms.

Section \ref{sec:setup} provides the set-up for our calculation.  We describe
the bulk geometry and use it to compute the geodesic distance between the
branes.  Doing this, it becomes clear that there is an issue in how to define
what we mean by the hierarchy of scales between the branes, which was the
point of the RS1 scenario \cite{rsa}.  We discuss this in some detail and
argue that when the bulk geometry has a horizon, the inter-brane separation
does {\it not\/} set the hierarchy, at least not in as direct a manner as when
the horizon is absent.

In section \ref{sec:cosmology} we solve the junction conditions and find the
cosmology on the branes.  We write down the bulk and boundary actions, the
equations of motion and the junction conditions, from which the cosmological
equations on both branes is derived.  We find as in \cite{langlois,csaki1},
that if the radion is not stabilized, there is a constraint between the energy
densities on the branes leading to one of them being negative.

The cosmology observed on a brane depends solely on its motion through the
bulk and since the two branes evolve independently there is no reason to
expect that for general densities on the branes they should both move in such
a way that the hierarchy is maintained.  In section \ref{sec:stablcosm} we
show how the presence of the Goldberger-Wise mechanism alters the picture.  In
addition to fixing the geodesic distance between the branes, the GW scalar
field introduces small time-dependent corrections to the bulk metric which
communicate the field content of each brane to the other.  These features can
be seen explicitly since the GW field can be treated as a small perturbation
to the AdS background.  We then consider an AdSS metric which contains a
horizon and find that the leading correction to the purely AdS cosmology is
suppressed compared with the corresponding result with only one brane
\cite{kraus,cosmologies}.

Section \ref{sec:conclusions} contains our conclusions.

\section{RS1 in AdSS Coordinates}\label{sec:setup}

Birkhoff's theorem in five dimensions tells us that the most general
spherically symmetric solution to the Einstein equations with a bulk
cosmological constant is given by the metric
\begin{equation}\label{eq:adsschwarzmetric}
    ds^2 = -u(r)\, dt^2 + {dr^2\over u(r)} + r^2\, d{\Sigma^2_k},
\end{equation}
where $r$ is the bulk coordinate, $d{\Sigma^2_k}$ is the line element for the
fixed $r$ 3-spatial hypersurfaces, which are homogeneous and isotropic, and
$k=0,\pm 1$ is the 3-spatial curvature parameter.  Writing the bulk
cosmological constant as $\Lambda=-6/l^2$, the metric coefficient $u(r)$ is
given by
\begin{equation}\label{eq:fk}
  u(r) \equiv {r^2\over l^2} + k - {\mu\over r^2},
\end{equation}
where $l$ is the AdS radius of curvature and $\mu$ is proportional to the mass
of the black hole in AdS \cite{db}.  In the RS2 picture this interpretation
can be made exact since there are asymptotic regions in $r$; in RS1, the
compactness in the $r$ direction prevents such a direct interpretation. 
However, for non-zero $\mu$, the solutions have horizons at $r=r_h$ where
\begin{equation}\label{eq:horizons}
  r^2_h= {l^2\over 2} \left( -k+\sqrt{k^2+4\mu/l^2} \right).
\end{equation}
In the RS2 case there is an explicit coordinate transformation
\cite{Ida,Maedaetal} that maps the above space-time to the cosmological metric
for RS2 as written in ref.~\cite{csaki1}.

\subsection{Geodesic Inter-brane Separation}\label{subsec:geodist}

The first question we need to ask is:  where is the radion mode?  It certainly
is {\it not\/} explicitly present in eq.~(\ref{eq:adsschwarzmetric}), in sharp
contrast to what happens when the RS1 scenario is worked out in Gaussian
normal coordinates \cite{gw,csaki1,gregoryetal}.  We can find this mode by
realizing that it is supposed to describe the inter-brane separation.  Given
eq.~(\ref{eq:adsschwarzmetric}), we can calculate this explicitly, by
computing the geodesic distance between points on the two branes.

Consider two branes placed at
\begin{eqnarray}
(t,r,x^i) &=& \left( T_0(\tau_0), R_0(\tau_0), x^i \right) , \nonumber \\
(t,r,x^i) &=& \left( T_1(\tau_1), R_1(\tau_1), x^i \right) ,
\label{eq:braneloc}
\end{eqnarray}
where the $\tau_{0,1}$ are the proper times on the branes.  We shall take
$R_0>R_1$ so that the brane at $R_1$ will be the TeV or IR brane while the
brane at $R_0$ is the Planck or UV brane.  The symmetry of the bulk metric
allows us to consider a geodesic between these branes that only propagates in
the $t$ and $r$ directions:
\begin{equation}\label{eq:geodesic}
  X^a(\lambda) = \left( t(\lambda), r(\lambda), 0, 0, 0 \right) ,
\end{equation}
where $\lambda$ is the affine parameter on the geodesic and we take
$\lambda=0,\ 1$ to correspond to the positions of the UV and IR brane
respectively.

The geodesic equations read
\begin{eqnarray}\label{eq:geodesiceqn}
  {d\over d\lambda} \left( u(r) {dt\over d\lambda} \right) &=& 0 \\
  {d\over d\lambda} \left( {1\over u(r)} {dr\over d\lambda} \right) +
{1\over 2} {\partial u\over\partial r} \left( {1\over u(r)} {dr\over d\lambda}
\right)^2 &=&
 - {1\over 2} {\partial u\over\partial r} \left( {dt\over d\lambda} \right)^2
. \nonumber
\end{eqnarray}
In the sequel, we shall frequently take $k=0$, {\it i.e.}~we take the fixed
$r$ hypersurfaces to have flat 3-spatial sections.  Note that in this case,
the horizon is located at $r_h=(\mu l^2)^{1/4}$.

The geodesic equations above have first integrals,
\begin{eqnarray}\label{eq:firstintegrals}
  {dt\over d\lambda} &=& {l\tilde L\over u(r(\lambda))}\nonumber \\
  {dr\over d\lambda} &=& l \sqrt{\tilde L^2 + \varepsilon^2
    u(r(\lambda)) } .
\end{eqnarray}
We have scaled all the dimensionful parameters by the AdS radius of curvature
$l$ so that $\tilde{L}$ and $\varepsilon$ are dimensionless.

We can solve this set of equations and then impose the boundary conditions
from eq.~(\ref{eq:braneloc}).  However, we can save some time by noting that
the relevant distance is the one measured at the {\it same\/} bulk time on
both branes.  This means that the geodesic we want is the one for which
$\tilde{L}=0$.  Integrating the $dr/d\lambda$ equation, we find
\begin{equation}\label{eq:roflambda}
    r^2(\lambda)=r_h^2 \cosh(2\varepsilon
    \lambda+\theta),
\end{equation}
with
\begin{equation}
\theta = {\cosh}^{-1} \left( {R_0^2\over r_h^2} \right)\ , \
2\varepsilon + \theta = {\cosh}^{-1} \left( {R_1^2\over r_h^2}
\right).
\end{equation}
Note that this assumes that both $R_0$, $R_1$ are larger than $r_h$.  The
equation for $r(\lambda)$ in eq.~(\ref{eq:firstintegrals}) shows that, with
$\tilde{L}=0$, the smallest value of $r$ allowed on this geodesic is $r_h$.

From eq.~(\ref{eq:firstintegrals}) we can compute the geodesic distance
between the branes:
\begin{eqnarray}\label{eq:geodistance}
  \Delta s &=& \int_0^1 d\lambda\, \sqrt{ \frac{\dot
r(\lambda)^2}{u(r(\lambda))}}
= |l \varepsilon|\nonumber\\
&=& {l\over 2} \left[ {\cosh}^{-1} \left( {R_0^2\over r_h^2} \right)
- {\cosh}^{-1} \left( {R_1^2\over r_h^2} \right) \right]. \qquad
\end{eqnarray}
In particular, in the $\mu\rightarrow 0$ ($r_h\to 0$) limit, we find:
\begin{equation}\label{RSGDrelation}
    \Delta s = -l \ln \frac{R_1}{R_0},\quad \text{or}\quad
    \frac{R_1}{R_0} = e^{-\Delta s/l}.
\end{equation}

\subsection{Hierarchy of Scales in AdSS Coordinates}\label{subsec:hierarchy}

We now ask how we see the fact that the separation between the branes gives
rise to a hierarchy of mass scales between the branes.

This is easy when  $\mu=0$, since $u(r)=r^2/l^2$ and the bulk line element can
be written as
\begin{equation}\label{nohorizonmetric}
    ds^2 = {r^2\over l^2} \left( - dt^2 + d\vec{x}^2 \right)
        + {l^2\over r^2}\, dr^2,
\end{equation}
where $d\vec{x}^2\equiv l^2 d\Sigma^2_{k=0}$.  If we now ask what sets the
scale on each brane, the fact that the branes (at least at a given instant of
bulk time) are located at a fixed value of $r$ allows us to set $r=R,\ dr=0$
which tells us that the relation between distance measurements on the two
branes is given by the ratio $R_1^2/R_0^2$.  As we saw above, this is just
$e^{-2\Delta s/l}$, which is exactly the standard RS result \cite{rsa}.

The $\mu \neq 0$ case contains some extra subtleties.  In this case, for $k=0$
the line element becomes
\begin{equation}\label{eq:horizonmetric}
   ds^2 = - \left( {r^2\over l^2} - {\mu\over r^2} \right)\, dt^2 
   + {r^2\over l^2}\, d\vec{x}^2
        + {dr^2\over {r^2\over l^2} - {\mu\over r^2}} .
\end{equation}
At fixed $r$ this metric will be {\it not\/} take the Minkowski form.  This
makes the comparison between branes somewhat less obvious, but we can ask how
distance scales on each brane compare in the case where all quantities are
{\it static\/} in bulk time $t$.  In this situation it is again the ratio
$R_1^2/R_0^2$ that sets the relative scales between the branes.  It is
important to note that this is {\it not\/} related to the inter-brane
separation in any simple way as can be seen from eq.~(\ref{eq:geodistance}).

This last point makes the issue of stabilization of the hierarchy somewhat
murkier in this formalism.  For the no-horizon case, stabilizing the
inter-brane distance stabilizes the hierarchy; with a horizon, there are two
variables to deal with, namely the ratio $R_1^2/R_0^2$ as well as one of the
positions $R_0$, say.  In the limit that $R_i^2\gg \sqrt{\mu l^2}=r_h^2$, we
recover the no-horizon results.

\section{RS1 Cosmology with an AdSS Bulk}\label{sec:cosmology}

The gravitational action for the RS1 scenario will be taken as the sum of the
bulk Einstein-Hilbert action with a cosmological constant $\Lambda$,
\begin{equation}\label{eq:RS1BulkAction}
  S_{\text{bulk}} = {1\over 16\pi G_5}\int d^5x\, \sqrt{-g}\,
 \left[ -2\Lambda + R \right],
\end{equation}
and a boundary action of the form:
\begin{widetext}
\begin{eqnarray}\label{eq:RS1BoundAction}
   S_{\text{brane}} &=&
{1\over 16\pi G_5}\int_0 d^4x\, \sqrt{-h_0}\,
   \left[ -2\sigma_0 + 16\pi G_5\, {\cal L}_0 \right]
+ {1\over 8\pi G_5} \int_0 d^4x\, \sqrt{-h_0}\, K_0
\nonumber\\
&+& {1\over 16\pi G_5}\int_1 d^4x\, \sqrt{-h_1}\,
\left[ -2\sigma_1 + 16\pi G_5\, {\cal L}_1 \right]
+ {1\over 8\pi G_5} \int_1 d^4x\, \sqrt{-h_1}\, K_1.
\end{eqnarray}
\end{widetext}
A subscript 0 (1) refers to the UV (IR) brane.  Here $G_5$ denotes the bulk
Newton constant, $h_{0,1}$, $K_{0,1}$ and $\sigma_{0,1}$ are the determinant
of the induced metric, the trace of the extrinsic curvature and brane surface
tension respectively for the appropriate brane.  If we define a unit normal
orthogonal to the tangent space of the branes by $[n_{0,1}]_a$, then the
induced metric and the extrinsic curvature are given in the bulk coordinates,
for example on the UV brane, by
\begin{eqnarray}
 [h_0]_{ab} &=& g_{ab} - [n_0]_a[n_0]_b \nonumber \\
 {[K_0]}_{ab} &=& [h_0]_a^{\ c} [h_0]_b^{\ d} \nabla_c [n_0]_d .
\label{eq:extrinsicdef}
\end{eqnarray}
The field content on each is summarized by ${\cal L}_{0,1}$; we shall
generally study the case where the fields on each brane produce the energy
momentum tensor of a perfect fluid.

Following \cite{kraus,cosmologies}, we allow the positions of the branes
within the bulk to evolve so as to give rise to cosmological evolution on each
brane.  The position of the brane in the direction transverse to the brane,
given in eq.~(\ref{eq:braneloc}), will be related to the cosmological scale
factor on the given brane. Now choose the normal to the brane to be of the
form
\begin{equation}\label{eq:normaltobrane}
    [n_0]_a = \pm \left( - \dot R_0, \dot T_0, 0, 0, 0 \right),
\end{equation}
where the dot denotes a derivative with respect to the proper time on the
appropriate brane. Noting that the condition
$g^{ab}n_an_b=1$ relates $\dot T_0$ to $\dot R_0$,
\begin{equation}\label{eq:Tdot}
    \dot T_0 = { \left[ \dot R_0^2 + u(R_0) \right]^{1/2}\over u(R_0)},
\end{equation}
we can write the induced metric in terms of brane coordinates
$(\tau_0, x^i)$:
\begin{eqnarray}\label{eq:inducedmetric}
 ds_0^2 &=& - u(R_0) \left[ \dot T_0^2 - (u(R_0))^{-2} \dot R_0^2 \right] \,
d\tau_0^2 + R_0^2(\tau_0)\, d\Sigma_k^2 \nonumber \\
&=& - d\tau_0^2 + R_0^2(\tau_0)\, d\Sigma_k^2
\nonumber \\
&\equiv& [h_0]_{\mu\nu}\, dx^\mu dx^\nu
\end{eqnarray}
where $\mu$, $\nu$ run over brane coordinates.  At the UV brane, the extrinsic
curvature becomes
\begin{eqnarray}\label{eq:extrinsiccurv}
[K_0]_{\mu\nu}\, dx^\mu dx^\nu
&=&\pm {1\over u(R_0(\tau_0)) \dot T_0}
\left[ \ddot R_0 + {1\over 2} {\partial u(R_0)\over \partial R_0} \right] \,
d\tau_0^2 \nonumber\\
& & \mp u(R_0) \dot T_0 R_0\, d\Sigma_k^2.
\end{eqnarray}

The signs that appear in the definition of the normal and consequently in the
extrinsic curvature, determine how the space is sliced.  We shall use the
upper sign in eq.~(\ref{eq:normaltobrane}) which corresponds to keeping the
space $r<R_0(\tau_0)$.  The form of the extrinsic curvature at the IR brane is
exactly the same, once we replace the subscripts, except that the lower signs
should be used to retain the region $r>R_1(\tau_1)$.

Let $p_0$ and $\rho_0$ be the pressure and density of the perfect fluid on the
UV brane so that the UV brane stress energy is
\begin{equation}\label{eq:UVstress}
    [T_0]_\mu^{\ \nu} = {\text{diag}}(-\rho_0, p_0, p_0, p_0).
\end{equation}

The Israel condition \cite{israel} relating the discontinuity in the extrinsic
curvature to the presence of the brane is given by
\begin{equation}\label{eq:israelUV}
\Delta [K_0]_{\mu\nu} = {1\over 3} \sigma_0 [h_0]_{\mu\nu}
+ 8\pi G_5 \left[ [T_0]_{\mu\nu}
- {1\over 3} [T_0]_\lambda^\lambda [h_0]_{\mu\nu} \right].
\end{equation}
We shall assume that the two $3$-branes reside at the fixed points of an
$S^1/\mathbb{Z}_2$ orbifold so that the jump in the extrinsic curvature is
$\Delta [K_0]_{\mu\nu} = 2 [K_0]_{\mu\nu}$.  Although eq.~(\ref{eq:israelUV})
appears to yield {\it two\/} constraints from the temporal and the spatial
components, the $\Delta [K_0]_{\tau\tau}$ constraint follows from the $\Delta
[K_0]_{ij}$ constraint \cite{kraus},
\begin{equation}\label{eq:spaceconstraint}
    \sqrt{\dot R_0^2+u(R_0)} = {1\over 6} R_0\sigma_0 +
\frac{4\pi G_5}{3} R_0\rho_0,
\end{equation}
provided energy and momentum are conserved on the brane
\begin{equation}\label{eq:conservation}
    \frac{d}{d\tau_0} \left( \rho_0 R_0^3 \right) = - p_0 \frac{d}{d\tau_0}
    R_0^3.
\end{equation}

The main advantage of letting the brane positions within the bulk evolve in
time is that the bulk metric is unaffected by the behavior of the branes other
than by their specifying which slice of the AdS-Schwarzschild metric
eq.~(\ref{eq:adsschwarzmetric}) is relevant.  Thus, for the UV brane we have
\begin{equation}\label{eq:UVcosm}
    \sqrt{\dot R_0^2+u(R_0)} = {1\over 6} R_0\sigma_0 +
    \frac{4\pi G_5}{3} R_0\rho_0 ,
\end{equation}
while on the IR brane,
\begin{equation}\label{eq:IRcosm}
    - \sqrt{\dot R_1^2+u(R_1)} = {1\over 6} R_1\sigma_1 + \frac{4\pi
G_5}{3} R_1\rho_1.
\end{equation}
The sign change corresponds to the fact that the normals of the two branes
have the opposite orientations.  As in ref.~\cite{kraus}, the evolution on the
UV brane approaches a standard Robertson-Walker cosmology when we make the
fine tuning $\sigma_0=6/l$ required in the original RS1 scenario and consider
the limit $R_0\gg l$.  On the IR brane, after squaring both sides of
eq.~(\ref{eq:IRcosm}), we see that we also need to impose $\sigma_1=\pm 6/l$
in order for the large terms in the $R_1\gg l$ limit to cancel,
\begin{eqnarray}\label{eq:IRlarge}
&&   \dot R_1^2 + \frac{R_1^2}{l^2} + k - \frac{\mu}{R_1^2} \\
&&\qquad\quad
= {1\over 36} R_1^2 \sigma_1^2 + {4\pi G_5\over 9} \sigma_1\rho_1 R_1^2
+ \left( {4\pi G_5\over 3} R_1\rho_1 \right)^2. \nonumber
\end{eqnarray}
In fact, we must choose $\sigma_1 = -6/l$ so as to satisfy the Israel
condition, eq.~(\ref{eq:IRcosm}), which in  turn, leads to a universe in which
the energy density $\rho_1$ must be negative to obtain standard FRW
cosmological evolution on the IR brane,
\begin{equation}\label{eq:FRWIR}
    \dot R_1^2 + k = - \frac{8\pi G_5}{3 l} \rho_1 R_1^2
+ \frac{\mu}{R_1^2} + \left( {4\pi G_5\over 3} \right)^2 \rho_1^2 R_1^2 .
\end{equation}

If the inter-brane separation is kept fixed at $\Delta z$ in the original
Gaussian normal coordinates of the RS model {\it without\/} a stabilization
mechanism at play, then it was shown in ref.~\cite{csaki1} that the energy
densities on the branes have to satisfy
\begin{equation}\label{eq:densityconstraint}
    \rho_0=-\rho_1 e^{-2 \Delta z/l}.
\end{equation}
How does this constraint emerge from the AdSS bulk coordinate approach? In
order to compare with ref.~\cite{csaki1} we set $\mu=0,\ k=0$, and impose the
fine tuning above: $\sigma_0 = -\sigma_1= 6/l$.  Now assume that the geodesic
distance between the branes has been fixed so that $\Delta s = -l
\ln(R_1/R_0)$ is constant. From this it follows that
\begin{eqnarray}\label{eq:scalefactor}
{1\over R_1} {dR_1\over d\tau_1}
   &=& {1\over R_0} {dR_0\over d\tau_1}
    = {d\tau_0\over d\tau_1} {1\over R_0} {dR_0\over d\tau_0} \nonumber \\
   &=& {\sqrt{ {R_0^2\over l^2} - {l^2\over R_0^2} {dR_0\over dt} }\over
        \sqrt{ {R_1^2\over l^2} - {l^2\over R_1^2} {dR_1\over dt} } }
      {1\over R_0} {dR_0\over d\tau_0} \nonumber \\
&\approx& {R_0\over R_1} {1\over R_0} {dR_0\over d\tau_0}
= e^{\Delta s/l} {1\over R_0} {dR_0\over d\tau_0}. \qquad
\end{eqnarray}
Using  eq.~(\ref{eq:scalefactor}) in the equations (\ref{eq:UVcosm}) and
(\ref{eq:IRcosm}), then 
\begin{equation}\label{UVIRtuned}
   \frac{\dot R_0^2}{R_0^2} = \frac{8\pi G_5}{3 l} \rho_0 + \cdots,
\quad
   \frac{\dot R_1^2}{R_1^2} = -\frac{8\pi G_5}{3 l} \rho_1 + \cdots,
\end{equation}
where the brane tensions satisfy the fine-tuning condition and we assume that
the energy densities are small compared to the tension.  In this limit, we
arrive at $\rho_0 \approx - \rho_1 e^{-2\Delta s/l}$.

\section{AdSS Cosmology with a Stabilization Mechanism}\label{sec:stablcosm}

When the bulk space-time only contains a cosmological constant, Birkhoff's
theorem allows us to write the metric in the conventional, time-independent
form given in eq.~(\ref{eq:adsschwarzmetric}).  The motion of the branes
through this bulk depends on the local environment of the branes, as expressed
by the Israel junction condition, so that without any means of interacting
each brane evolves independently.  However, the existence of a fixed hierarchy
requires that the motions of the branes must be carefully correlated.  From
this perspective then, the appearance of an unphysical energy density is not
altogether surprising.  We have simply imposed a constraint upon two
independently moving branes without any mechanism to enforce it.

To evade Birkhoff's theorem, the mechanism that stabilizes the hierarchy must
distort the background away from a pure AdSS space-time.  These distortions
will then appear in the extrinsic curvature term of the Israel equation which
thus allows the motions of the branes to be naturally correlated with real,
positive, independent densities on each brane.  Because of this correlated
motion of the branes, we generally expect that the cosmology observed on
either of the branes now should depend on the field content of both, as was
seen in \cite{csaki1}.  Note that this distortion can be small compared with
the scale of the bulk cosmological constant---thus allowing a perturbative
treatment---since the terms that yielded an FRW cosmology on the branes were
subleading and only became important once the tension was finely tuned to
cancel the leading bulk effect.

The Goldberger-Wise stabilization mechanism \cite{gw} adds a free massive
scalar field to the bulk with a potential on each brane to fix the boundary
values of the scalar field,
\begin{eqnarray}
S_{\rm GW} &=& {1\over 8\pi G_5} \int d^5x\, \sqrt{-g}\, \left[ -\frac{1}{2}
\nabla_a\phi\nabla^a\phi - \frac{1}{2} m^2 \phi^2
\right] \nonumber \\
&& + {1\over 8\pi G_5} \int_0 d^4x\, \sqrt{-h_0}\, \left[ -\lambda_0
(\phi^2-v_0^2)^2 \right]
\label{eq:gwaction} \\
&& + {1\over 8\pi G_5} \int_1 d^4x\, \sqrt{-h_1}\, \left[ -\lambda_1
(\phi^2-v_1^2)^2 \right] .
\nonumber
\end{eqnarray}
We have normalized the fields to extract the factor $(8\pi G_5)^{-1}$ to
simplify the form of some of the later equations.  Varying the total action
produces the usual Einstein and scalar field
equations in the bulk,
\begin{eqnarray}
 R_{ab} - {\textstyle{1\over 2}} g_{ab} R &=& - \Lambda g_{ab} + {\cal T}_{ab}
\nonumber \\
 \nabla^2\phi + m^2 \phi &=& 0 ,
% \Box\phi - m^2 \phi &=& 0 ,
\label{eq:bulkeom}
\end{eqnarray}
while at the UV brane the equations of motion are
\begin{eqnarray}
  \Delta [K_0]_{ab} &=& {\textstyle{1\over 3}} \sigma [h_0]_{ab}
  + 8\pi G_5 \left[ [T_0]_{ab} - {\textstyle{1\over 3}} [T_0]_c^c
  [h_0]_{ab} \right] \nonumber \\
 {[n_0]}^a\partial_a\phi &=& 2\lambda_0 (\phi^2-v_0^2) \phi
\label{eq:braneeom}
\end{eqnarray}
with an analogous equation at the IR brane. Here, $[T_0]_{ab}$ and ${\cal
T}_{ab}$
are the energy-momentum tensors associated with the fields confined to the
brane and with $\phi$ respectively.

The presence of the scalar field alters the bulk geometry, but in the limit
where it is small compared to the cosmological constant, $ml\ll 1$, we can
treat its effect as a perturbation to the AdSS background,
\begin{eqnarray}
  ds^2 &=& -u(r) \left[ 1 + \chi_{tt}(t,r) \right]\, dt^2
\label{eq:perturbmetric} \\
&& + {1\over u(r)} \left[ 1 + \chi_{rr}(t,r) \right]\, dr^2
  + r^2\, d\Sigma^2_{k=0},
\nonumber
\end{eqnarray}
where throughout this section we set $k=0$.  In this metric we have still
assumed that the three large spatial dimensions are isotropic.  Since the
scalar field is responsible for the perturbations, $\chi_{tt}$ and $\chi_{rr}$
will be of the same order as ${\cal T}_{ab}$.

As the induced metric at the brane suggests, the scale $R_1$ ($< R_0$) is
associated with the scale factor of an FRW universe while $l$ is naturally of
the order of the Planck length.  Therefore, for an acceptable cosmology, we
can assume that $l/r\ll 1$ throughout the bulk.  In this limit the form of the
metric naturally suppresses terms with time derivatives relative to those with 
$r$ derivatives.  This feature greatly simplifies the analysis of the
back-reaction of the scalar field on the bulk geometry since
eq.~(\ref{eq:bulkeom}) will effectively only constrain the $r$ dependence of
$\chi_{tt}$, $\chi_{rr}$ and $\phi$. This behavior can be seen, for example,
by considering ${\cal T}_{tt}$ which is given up to $\chi_{tt}$, $\chi_{rr}$
corrections by
\begin{equation}
{\cal T}_{tt} = {\textstyle{1\over 2}} (\partial_t\phi)^2
+ {\textstyle{1\over 2}} u^2(r) (\phi')^2
+ {\textstyle{1\over 2}} u(r) m^2 \phi^2 + \cdots
\label{eq:Ttteg}
\end{equation}
with $\phi'\equiv \partial_r\phi$.  If neither brane is near the black hole
horizon, then the $r$-derivative term is enhanced by a factor of
$u^2(r)\approx r^4/l^4$ relative to the time derivative term.  The same
feature appears in the Einstein equations eq.~(\ref{eq:bulkeom}).  Thus, as
long as the time derivatives are not excessively large, as we shall later
show, the bulk field equations only constrain the radial dependence.

Expanding the field equations to first order in $\chi_{tt}$, $\chi_{rr}$ and
$\phi^2$, and substituting in the zeroth order solution for an AdS (or AdSS)
background, we find
\begin{eqnarray}
&&{12\over l^2} \chi_{rr} + {3u\over r} \chi'_{rr}
= u {\phi'}^2 + m^2 \phi^2 \nonumber \\
&&{12\over l^2} \chi_{rr} - {3u\over r} \chi'_{tt}
= - u {\phi'}^2 + m^2 \phi^2 \nonumber \\
&&{3\over r} \partial_t\chi_{rr} = 2\phi' \partial_t\phi \label{eq:gfirst} \\
&&{12\over l^2} \chi_{rr}
+ \left[ 2{r^2\over l^2} + u \right] {\chi'_{rr}\over r}
- \left[ 6{r^2\over l^2} - u \right] {\chi'_{tt}\over r} \nonumber \\
&&\qquad
- u \chi_{tt}^{\prime\prime}
= u {\phi'}^2 + m^2 \phi^2
\nonumber
\end{eqnarray}
from the Einstein equations and
\begin{equation}
  u\phi^{\prime\prime} + u'\phi' + {3u\over r}\phi' - m^2\phi = 0
\label{eq:phifirst}
\end{equation}
from the scalar field equation (\ref{eq:bulkeom}).  The Bianchi identity
relates these equations so that, for example, the last equation in
(\ref{eq:gfirst}) is not independent.

\subsection{A Stabilized AdS Cosmology}\label{AdSstabilized}

For a purely AdS space, {\it i.e.\/}~$\mu=0$, and in the limit that we
suppress the contributions of time derivatives, we can solve the scalar field
solution as in to find as in ref.~\cite{gw},
\begin{equation}
\phi(r) = {1\over r^2} \left[ a(t) r^\nu + b(t) r^{-\nu} \right] ,
\label{eq:AdSphifirst}
\end{equation}
where
\begin{equation}
\nu\equiv\sqrt{4+m^2 l^2} .
\label{eq:nudef}
\end{equation}
Substituting this solution into eq.~(\ref{eq:gfirst}) and solving for the
metric perturbations yields
\begin{eqnarray}
\chi_{tt}(t,r) &=& {1\over 3} {a(t)b(t)(\nu^2-4) - 3c(t)\over r^4} + d(t)
\nonumber \\
\chi_{rr}(t,r) &=& {1\over 3} a^2(t) (\nu-2) r^{-2(2-\nu)} \label{eq:ABsoln}
\\
& &
- {1\over 3} b^2(t) (\nu+2) r^{-2(2+\nu)} + {c(t)\over r^4} . \quad\qquad
\nonumber
\end{eqnarray}
The functions $a(t)$, $b(t)$, $c(t)$, and $d(t)$ are constants of integration
with respect to the $r$ derivatives and are only fixed to this order in $l/r$
by the third equation of (\ref{eq:gfirst}) which requires that
\begin{equation}
\partial_t c = - {\textstyle{2\over 3}} (\nu+2) b \partial_t a
+ {\textstyle{2\over 3}} (\nu-2) a\partial_t b .
\label{tconstraint}
\end{equation}

The scalar field must satisfy the jump conditions described by
eq.~(\ref{eq:braneeom}).  However, if as in \cite{gw} we assume that the
potentials on the branes are sufficiently rigid,
$\lambda_0,\lambda_1\to\infty$, then the scalar field value is forced to the
minimum value of the potential on each brane,
\begin{eqnarray}
\phi(T_0,R_0) &=& v_0 = a(T_0) R_0^{-2+\nu} + b(T_0) R_0^{-2-\nu} \nonumber \\
\phi(T_1,R_1) &=& v_1 = a(T_1) R_1^{-2+\nu} + b(T_1) R_1^{-2-\nu} . \qquad
\label{eq:phiconstraint}
\end{eqnarray}
Now, we saw in section \ref{sec:setup} that the stabilized hierarchy in these
coordinates corresponds to fixing the geodesic distance between the branes
along a {\it constant time\/} geodesic, so that $T_0=T_1$.  Setting
$a=a(T_0)=a(T_1)$ and $b=b(T_0)=b(T_1)$, we thus find
\begin{eqnarray}
a &=& {v_0 - v_1 (R_1/R_0)^{2+\nu} \over 1 - (R_1/R_0)^{2\nu} } R_0^{2-\nu}
\nonumber \\
b &=& {v_1 (R_1/R_0)^{2-\nu} - v_0\over 1 - (R_1/R_0)^{2\nu} } \left(
{R_1\over R_0} \right)^{2\nu}  R_0^{2+\nu} .
\label{eq:absolns}
\end{eqnarray}
when evaluated at either brane.  This result is exactly that of ref.~\cite{gw}
except for the implicit time dependence in $R_0$ and $R_1$.

When the scalar field satisfies eq.~(\ref{eq:phiconstraint}), the scalar
potentials vanish and the only contribution to the energy-momentum tensor on
the branes is due to the fields confined to the branes which we take to be a
perfect fluid stress tensor, as in  eq.~(\ref{eq:UVstress}).  For fluids
satisfying the conservation law of eq.~(\ref{eq:conservation}), we only need
to solve for the $ij$ component since the $\tau\tau$ component of the Israel
condition is does not give rise to an independent constraint.  Retaining only
the corrections linear in
the perturbations and unsuppressed by powers of $l/r$ in
\begin{equation}
[K_0]_{ij} = {\sqrt{u(R_0) + \dot R_0^2}\over R_0} \left[ 1 -
{\textstyle{1\over 2}} \chi_{rr}(T_0,R_0) + \cdots \right] [h_0]_{ij},
\label{Kij}
\end{equation}
the Israel condition on the UV brane becomes
\begin{equation}
2 {R_0^2\over l^2} + \dot R_0^2 = {R_0^2\over 3l}\sigma_0 + {8\pi G_5\over 3l}
\rho_0 R_0^2 + {R_0^2\over l^2}\chi_{rr}(T_0,R_0).
\label{eq:israelijUV}
\end{equation}
At the IR brane, since we have different signs from the opposite orientation
for the normal, we arrive at 
\begin{equation}
2 {R_1^2\over l^2} + \dot R_1^2 = - {R_1^2\over 3l}\sigma_1 - {8\pi G_5\over
3l} \rho_1 R_1^2 + {R_1^2\over l^2}\chi_{rr}(T_1,R_1) .
\label{eq:israelijIR}
\end{equation}
In both of these equations, we have neglected ${\cal O}(\rho^2)$ corrections.

The small change in the bulk background produced by the Goldberger-Wise field
will in general require a corresponding shift in the usual choice of the brane
tensions to cancel the cosmological constant in the effective cosmology on the
brane.  We therefore set the brane tensions to:
\begin{equation}
\sigma_0 = {6\over l} \left( 1 + {1\over 2} \delta\sigma_0 \right)
\quad\hbox{and}\quad
\sigma_1 = - {6\over l} \left( 1 + {1\over 2} \delta\sigma_1 \right)
\label{branetensions}
\end{equation}
Substituting these into (\ref{eq:israelijUV}) and (\ref{eq:israelijIR}) and
using our solution for the metric perturbation in (\ref{eq:ABsoln}) yields
\begin{eqnarray}
\dot R_0^2 &=& {8\pi G_5\over 3l} \rho_0 R_0^2
+ {c(T_0)\over l^2} {1\over R_0^2} \nonumber \\
&&
+ \left[ \delta\sigma_0 - {4(v_1-v_0)^2\over 3} {R_1^8\over R_0^8} \right]
{R_0^2\over l^2} + \cdots \nonumber \\
\dot R_1^2 &=& - {8\pi G_5\over 3l} \rho_1 R_1^2
+ {c(T_1)\over l^2} {1\over R_1^2} \nonumber \\
&&
+ \left[ \delta\sigma_1 - {4(v_1-v_0)^2\over 3} \right] {R_1^2\over l^2} +
\cdots . \label{eq:leadcosmoonbranes}
\end{eqnarray}
Since we wish to establish that the leading behavior mimics a FRW cosmology,
we have neglected ${\cal O}(ml)$ corrections in (\ref{eq:leadcosmoonbranes})
by setting $\nu=2$.

The final function $c(t)$ is determined on the branes by the requirement that
the hierarchy should be stabilized.  As in \cite{gw}, we can determine the
preferred separation between the branes for a given set of parameters $(m,
v_0, v_1)$ by integrating the scalar action over the radial dimension to
define an effective potential for $R_1/R_0$,
\begin{eqnarray}
& &\int d^4x\, \sqrt{-g_{\rm eff}} \left[ - V_{\rm eff}(R_0,R_1) \right]
\label{eq:Veffdef} \\
& & \qquad
\equiv
\int_{R_1}^{R_0} dr \int d^4x\, \sqrt{-g} \left[ -\frac{1}{2}
\nabla_a\phi\nabla^a\phi -\frac{1}{2}
m^2 \phi^2 \right] .
\nonumber
\end{eqnarray}
Substituting eq.~(\ref{eq:AdSphifirst}) into eq.~(\ref{eq:Veffdef}) and
integrating over a constant time hypersurface so that we can use
eq.~(\ref{eq:absolns}), we discover that the dependence on geodesic distance
$R_1/R_0$ in the effective potential factorizes
\begin{equation}
V_{\rm eff}(R_0,R_1) = {R_0^4\over l^2} \hat V_{\rm eff}\left( {R_1\over R_0}
\right)
\label{eq:Vefffactor}
\end{equation}
with
\begin{eqnarray}
\hat V_{\rm eff}(x) &=&
{ (\nu-2) \left[ v_0 - v_1\, x^{2+\nu} \right]^2 \over 1 - x^{2\nu}}
\nonumber \\
& &
+ { (\nu+2) \left[ v_0 - v_1\, x^{2-\nu} \right]^2  x^{2\nu} \over 1 -
x^{2\nu}}
. \qquad \label{eq:Veffsoln}
\end{eqnarray}
In the $ml\ll 1$ limit, $\hat V_{\rm eff}(x)$ is minimized by
\begin{equation}
x = {R_1\over R_0} \approx \left( {v_1\over v_0} \right)^{4/m^2l^2} .
\label{eq:Veffminimum}
\end{equation}
Note that by adding an $\mathcal{O}(m^2l^2)$ constant to the right side of
eq.~(\ref{eq:Veffdef}), we can arrange the minimum to occur at $\hat V_{\rm
eff}=0$, thus keeping the $R_0^4$ coefficient in eq.~(\ref{eq:Vefffactor})
from driving $R_0\to 0$ or $R_0\to\infty$.

Once the relative brane separation has settled to the value determined by
eq.~(\ref{eq:Veffminimum}), the motion of the branes is correlated so that, as
in eq.~(\ref{eq:scalefactor}),
\begin{equation}
\dot R_0 = {dR_0\over d\tau_0} \approx \dot R_1 = {dR_1\over d\tau_1} .
\label{eq:AdSgeodot}
\end{equation}
In the absence of a stabilizing mechanism, this condition was arbitrarily
imposed as a constraint on the two brane system which required a negative
energy density on one of the branes.  We now see that eq.~(\ref{eq:AdSgeodot})
arises as a natural consequence of minimizing the radion effective potential
and instead of constraining the field densities on the branes, it fixes the
remaining integration function $c(t)$ evaluated on the branes at equal times,
$c\equiv c(T_0) = c(T_1)$,
\begin{eqnarray}
c &=& l^2 R_1^2 \left( 1 - {R_1^2\over R_0^2} \right)^{-1} \biggl[
{8\pi G_5\over 3l} (\rho_0 R_0^2 + \rho_1 R_1^2) \label{eq:csoln} \\
& &
+ {1\over l^2} \left( \delta\sigma_0 R_0^2 - \delta\sigma_1 R_1^2 \right)
+ {4\over 3l^2} (v_1-v_0)^2 R_1^2 \biggr] . \qquad
\nonumber
\end{eqnarray}
When eq.~(\ref{eq:csoln}) is substituted into the equation for the evolution
of the IR brane, the leading $\rho_1$ term cancels the density term in
(\ref{eq:leadcosmoonbranes}) with the undesired sign.  The next to leading
term in $\rho_1$, where the expansion is in powers of the exponential
hierarchy, has the required sign for a well-behaved FRW cosmology:
\begin{eqnarray}
\dot R_1^2 &=& {8\pi G_5\over 3l} {R_1^2\over R_0^2} \left[
\rho_1 + \rho_0 {R_0^4\over R_1^4} \right] R_1^2 \label{eq:IRcosmology} \\
& &
+ \left[ \delta\sigma_0 {R_0^4\over R_1^4} - \delta\sigma_1
+ {4\over 3l^2}  (v_1-v_0)^2 \right] {R_1^2\over R_0^2} R_1^2 + \cdots .
\nonumber
\end{eqnarray}
Choosing the tensions such that the second term vanishes, we see that the
presence of a stabilized radion has led to a realistic effective cosmology on
the branes,
\begin{equation}
{\dot R_1^2\over R_1^2} = {8\pi G_5\over 3l} e^{-2\Delta s/l} \left[
\rho_1 + \rho_0 e^{4\Delta s/l} \right]
+ \cdots .
\label{eq:IRtunedcos}
\end{equation}
The fixed hierarchy implies that the equation for the evolution of the UV
brane is exactly the same up to a rescaling by $e^{-2\Delta s/l}$,
\begin{equation}
{\dot R_0^2\over R_0^2} = {8\pi G_5\over 3l}
\left[ \rho_0 + \rho_1 e^{-4\Delta s/l}\right]
+ \cdots .
\label{eq:UVtunedcos}
\end{equation}

We might have na\"{\i}vely expected that once the hierarchy between the two
brane system is stabilized, the cosmological evolution should depend on the
field content of both branes, weighted by possible exponential factors.  The
approach where the brane cosmology arises from the motion of the branes
through a nearly static bulk explicitly shows the origin of this dependence. 
By distorting the bulk from pure anti-de Sitter space, the scalar field both
communicates the energy density from one brane to the other as in
eq.~(\ref{eq:csoln}) and cancels the unphysical term in the IR brane motion. 
As noted in \cite{csaki1}, the presence of matter on the UV brane can easily
overwhelm the effect of matter on the IR brane and so drive the cosmology
unless it is exponentially smaller than that of the IR brane.  Also,
eq.~(\ref{eq:IRtunedcos}) suggests that we should define the effective Newton
constant for an observer on the IR brane to be
\begin{equation}
G_N \equiv {G_5\over l} e^{-2\Delta s/l} ;
\label{eq:effNewton}
\end{equation}
this result also agrees with the effective Newton constant appropriate for two
masses confined to the IR brane, as derived in \cite{csaki1}.

We can also discuss some of the corrections to eq.~(\ref{eq:IRtunedcos}) which
would allow us to distinguish a brane-induced cosmology from a standard FRW
cosmology.  As in eq.~(\ref{eq:FRWIR}), the presence of $\rho_1^2$ effects
reflects a general feature of brane-induced cosmologies.  To find the detailed
form of such corrections would require solving the equations of motion to
second order in the metric perturbations since, from
eqs.~(\ref{eq:csoln},\ref{eq:ABsoln}), $\chi_{rr}$ has terms that depend
linearly on the field densities of the branes.  The leading $\mathcal{O}(ml)$
corrections are proportional to $R_1^2$, or $R_0^2$, and can be eliminated by
appropriately choosing the brane tensions.  If we do not fine-tune the brane
tensions, then the cosmology will contain a term resembling an effective
cosmological constant.

The solutions to the perturbed metric were found in the limit in which the
time derivatives are suppressed in the equations of motion.  We can determine
when these terms would be important, thus requiring that the full $r$ and $t$
dependent equations be solved, by comparing the typical size of of such terms
in the solutions presented above.  For example, the scalar potentials in
eq.~(\ref{eq:gwaction}) hold the value of the field constant at the branes so
that as the brane moves, the $r$ and $t$ derivatives are related via
\begin{equation}
\dot T_0 \partial_t\phi + \dot R_0 \phi' = {d\phi\over d\tau_0} = 0.
\label{eq:dphidrdt}
\end{equation}
Thus, using eq.~(\ref{eq:Tdot}) and neglecting $\dot R_0^2$ compared to
$u(R_0)=R_0^2/l^2$,
\begin{equation}
\partial_t\phi \approx - {R_0\over l} \dot R_0 \phi' .
\label{eq:dphidtatbrane}
\end{equation}
As we saw earlier in eq.~(\ref{eq:Ttteg}), the $\phi'$ terms are enhanced by a
factor $u^2(r)=r^4/l^4$ over $\partial_t\phi$ terms,
\begin{eqnarray}
{\cal T}_{tt} &=& {\textstyle{1\over 2}}(\partial_t\phi)^2 +
{\textstyle{1\over 2}} u^2(R_0) (\phi')^2 + \cdots \nonumber \\
&=& {1\over 2} {R_0^4\over l^4} \left[ {l^2\over R_0^2} \dot R_0^2(\phi')^2
+ (\phi')^2 \right] + \cdots ;
\label{eq:Tttatbrane}
\end{eqnarray}
so that despite eq.~(\ref{eq:dphidtatbrane}), the first term is still
negligible compared with the second.

\subsection{A Stabilized AdSS Cosmology}\label{stabilized}

The meaning of a stabilized hierarchy becomes less clear when a black hole
horizon is introduced into the bulk metric, even if the horizon does not
actually appear in the bulk.  The parameter $\mu$ in the anti-de
Sitter-Schwarzschild metric breaks the conformal flatness of the metric in the
large $3+1$-dimensions, which is needed to define an unambiguous hierarchy. 
Yet when this parameter is sufficiently small
\begin{equation}
{r^2\over l^2} - {\mu\over r^2} \gg 1
\qquad\quad\hbox{for $R_1 \le r \le R_0$}
\label{eq:AdSSineq}
\end{equation}
so that we can still neglect the time derivatives, the quantity that sets the
hierarchy and therefore must be fixed by the stabilization mechanism is still
$R_1/R_0\equiv e^{-\Delta s/l}$, up to $\mu$-dependent corrections.  In fact,
if these corrections are to be larger than those due to the time derivatives
that we are neglecting, then the black hole mass should not be too small,
\begin{equation}
1 \gg {l\sqrt{\mu}\over r^2} , {l\over\sqrt{\mu} } .
\label{eq:horizonpos}
\end{equation}

Neglecting time derivatives again, the important terms in the bulk Einstein
equations
eq.~(\ref{eq:gfirst}) and bulk scalar field equation eq.~(\ref{eq:phifirst})
are
\begin{eqnarray}\label{eq:AdSSfirst}
12\chi_{rr} + {3\over r} {r^4-\mu l^2\over r^2} \chi'_{rr}
&=& {r^4-\mu l^2\over r^2} {\phi'}^2 + m^2l^2 \phi^2 \qquad\  \\
12\chi_{rr} - {3\over r} {r^4-\mu l^2\over r^2} \chi'_{tt}
&=& - {r^4-\mu l^2\over r^2} {\phi'}^2 + m^2l^2 \phi^2 \nonumber 
\end{eqnarray}
and
\begin{equation}
{r^4-\mu l^2\over r^2}\phi^{\prime\prime} + {5r^4-\mu l^2\over r^3}\phi'
- m^2l^2\phi + \cdots = 0 .
\label{eq:AdSSleadingphi}
\end{equation}

While we require $ml>0$ in order to determine the effect of the presence of
the horizon on the Goldberger-Wise mechanism, once the hierarchy has been
stabilized, ${\cal O}(ml)$ terms will be unimportant for the leading
description of the brane cosmology.  Setting $ml\to 0$ in
eqs.~(\ref{eq:AdSSfirst}--\ref{eq:AdSSleadingphi}, yields the following
leading behavior for the scalar field,
\begin{equation}
\phi(r) = a(t) - {b(t)\over\mu l^2} \ln \left[ 1 - {\mu l^2\over r^4} \right]
+ {\cal O}(ml)
\label{AdSSphinom}
\end{equation}
while for corrections to the background metric we find,
\begin{eqnarray}\label{eq:AdSSABsolnnom}
\chi_{tt}(t,r) &=& d(t) + {-1\over r^4-\mu l^2} \biggl[ c(t) + {8\over
3}{b^2(t)\over
\mu l^2} \\
& &
+ {8\over 3}{b^2(t) r^4\over\mu^2 l^4} \left(1 - {1\over 2}{\mu l^2\over r^4}
\right)
\ln \left[ 1 - {\mu l^2\over r^4} \right] \biggr]
 \nonumber \\
\chi_{rr}(t,r) &=& {4\over 3} {b^2(t)\over\mu l^2(r^4-\mu l^2)} \ln \left[ 1 -
{\mu l^2\over r^4} \right] + {c(t)\over r^4-\mu l^2},
\quad \nonumber 
\end{eqnarray}
up to ${\cal O}(ml)$ terms.

Determining the boundary conditions as before and retaining only the leading
$\mu$-dependent corrections yields the following cosmology on the IR brane
\begin{eqnarray}
\dot R_1^2
&=& {8\pi G \over 3l} {R_1^2\over R_0^2} \left[
\rho_1 + {R_0^4\over R_1^4} \rho_0 \right] R_1^2
+ {\cal O}(\mu^2)
\label{eq:AdSSleading} \\
& &\quad
+ \left[ {R_0^4\over R_1^4} \delta\sigma_0 - \delta\sigma_1
+ {4\over 3} (v_1-v_0)^2 \right] {R_1^2\over R_0^2} {R_1^2\over l^2} .
\qquad \nonumber
\end{eqnarray}
Upon fine-tuning the brane tensions,
\begin{equation}
{\dot R_1^2\over R_1^2} = {8\pi G_N \over 3}
\left[ \rho_1 + \rho_0 e^{4\Delta s/l} \right]
+ {\cal O}(\mu^2) + \cdots ,
\label{eq:AdSSfinetuned}
\end{equation}
where $e^{-\Delta s/l}\equiv R_1/R_0$ as before and the effective Newton
constant is defined in (\ref{eq:effNewton}).  Notice that unlike the scenario
with a single brane \cite{kraus,cosmologies}, where the leading terms in the
brane cosmology are of the form
\begin{equation}
{\dot R^2\over R^2} = {8\pi G_5 \over 3l} \rho + {\mu\over R^4} + \cdots ,
\label{eq:onebrane}
\end{equation}
such a term, linear in $\mu$, does not appear in the two brane scenario.  In
the one brane scenario, such a term behaves, with its $R^{-4}$ dependence,
like a radiation fluid but is canceled here by the stabilization mechanism. 
Therefore, the distance between the IR brane and the horizon can be much
smaller than in the single brane model and still produce an acceptable
cosmology, at least until terms quadratic in $\rho_{0,1}$ and $\mu$ become
important.

\section{Conclusions}\label{sec:conclusions}

There were two main points we wanted to make in this work.  The first was
that, while some of the features of RS1 models are less obvious in the AdSS
formulation presented above, and despite the fact that the bulk is static to
leading order in this formulation, when the black hole horizon vanishes, the
brane cosmologies have the same behavior here as in the Gaussian normal
formulation.  This was not {\it a priori\/} obvious, especially because the
bulk geometry appears quite different in the two different formulations.

The second point was that introducing a horizon into the geometry can modify
the cosmological behavior on the IR brane.  While we were only able to see
this in a perturbative expansion in $r_h\slash r$, we were able to see that
the stabilization mechanism cancels the leading effect so that the ``dark
radiation'' contribution that generically appears in the one brane scenario is
absent her.

It would be interesting to go beyond a perturbative solution in the $\mu\neq
0$ case; this is technically complicated and may involve some conceptual
issues such as how the horizon affects the boundary conditions on bulk fields,
the GW scalar in particular.  We hope to return to this in a future
publication.

\begin{acknowledgments}
We thank Ira Rothstein for suggesting this problem to us and for useful
discussions at various points in this work.  This work was supported in part
by DOE grant DE-FG03-91-ER40682.
\end{acknowledgments}


\begin{thebibliography}{99}
\bibitem{rsa}
L.~Randall and R.~Sundrum,
%``A large mass hierarchy from a small extra dimension,''
Phys.\ Rev.\ Lett.\  {\bf 83}, 3370 (1999) [hep-ph/9905221].

\bibitem{rsb}
L.~Randall and R.~Sundrum,
%``An alternative to compactification,''
Phys.\ Rev.\ Lett.\  {\bf 83}, 4690 (1999) [hep-th/9906064].

\bibitem{kraus}
P.~Kraus,
%``Dynamics of anti-de Sitter domain walls,''
JHEP {\bf 9912}, 011 (1999) [hep-th/9910149].

\bibitem{cosmologies}
H.~Collins and B.~Holdom,
%``Brane cosmologies without orbifolds,''
Phys.\ Rev.\ D {\bf 62}, 105009 (2000) [hep-ph/0003173].

\bibitem{chambreall}
H.~A.~Chamblin and H.~S.~Reall,
%``Dynamic dilatonic domain walls,''
Nucl.\ Phys.\ B {\bf 562}, 133 (1999) [hep-th/9903225].

\bibitem{others}
See for example: S.~Davis,
%``Brane Cosmology Solutions with Bulk Scalar Fields,''
(2001) [hep-ph/0111351] ;
Y.~S.~Myung,
%``Quintessence with a Localized Scalar Field on the Brane,''
Mod.\ Phys.\ Lett.\ A {\bf 16} 1963 (2001) [hep-th/0108092].

\bibitem{israel}
W.~Israel,
%``Singular Hypersurfaces And Thin Shells In General Relativity,''
Nuovo Cim.\ B {\bf 44S10}, 1 (1966) [Erratum-ibid.\ B {\bf 48}, 463 (1966)].

\bibitem{adscft}
J.~Maldacena,
%``The large $N$ limit of superconformal field theories and supergravity,''
Adv.\ Theor.\ Math.\ Phys.\  {\bf 2}, 231 (1998)
[Int.\ J.\ Theor.\ Phys.\  {\bf 38}, 1113 (1998)]
[hep-th/9711200].

\bibitem{csaki1}
C.~Csaki, M.~Graesser, L.~J.~Randall and J.~Terning,
%``Cosmology of brane models with radion stabilization,''
Phys.\ Rev.\ D {\bf 62}, 045015 (2000) [hep-ph/9911406].

\bibitem{csaki2}
C.~Csaki, M.~Graesser, C.~Kolda and J.~Terning,
%``Cosmology of one extra dimension with localized gravity,''
Phys.\ Lett.\ B {\bf 462}, 34 (1999) [hep-ph/9906513].

\bibitem{cline}
J.~M.~Cline, C.~Grojean and G.~Servant,
%``Cosmological expansion in the presence of extra dimensions,''
Phys.\ Rev.\ Lett.\  {\bf 83}, 4245 (1999) [hep-ph/9906523].

\bibitem{ira}
I.~Z.~Rothstein,
%``Cosmological solutions on compactified AdS(5) with a thermal bulk,''
Phys.\ Rev.\ D {\bf 64}, 084024 (2001) [hep-th/0106022].

\bibitem{langlois}
P.~Binetruy, C.~Deffayet, U.~Ellwanger, and D.~Langlois,
%``Brane cosmological evolution in a bulk with cosmological constant,''
Phys.\ Lett.\ B {\bf 477}, 285 (2000) [hep-th/9910219].

\bibitem{gw}
W.~D.~Goldberger and M.~B.~Wise,
%``Modulus stabilization with bulk fields,''
Phys.\ Rev.\ Lett.\  {\bf 83}, 4922 (1999) [hep-ph/9907447].

\bibitem{holo}
R.~Rattazzi and A.~Zaffaroni,
%``Comments on the Holographic Picture of the Randall-Sundrum Model''
JHEP 0104:021,(2001), [hep-th/0012248];
Nima Arkani-Hamed, Massimo Porratti and Lisa Randall
%``Holography and Phenomenology''
JHEP {\bf 0108}, 017, (2001) [hep-th/0012148] ;
Paolo Creminelli, Alberto Nicolis and Ricardo Rattazzi
%``Holography and the Electroweak Phase Transition''
, (2001) [hep-th/0107141];
Tetsuya Shiromizu, Takashi Torii, Daisuke Ida
%``Brane World and Holography''
(2001) [hep-th/0105256] .

\bibitem{db}
D.~Birmingham,
%``Topological black holes in anti-de Sitter space,''
Class.\ Quant.\ Grav.\  {\bf 16}, 1197 (1999) [hep-th/9808032].

\bibitem{Ida}
D.~Ida,
%``Brane-world cosmology,''
JHEP {\bf 0009}, 014 (2000) [gr-qc/9912002].

\bibitem{Maedaetal}
S.~Mukohyama, T.~Shiromizu and K.~i.~Maeda,
%``Global structure of exact cosmological solutions in the brane world,''
Phys.\ Rev.\ D {\bf 62}, 024028 (2000)
[Erratum-ibid.\ D {\bf 63}, 029901 (2000)] [hep-th/9912287].

\bibitem{gregoryetal}
C.~Charmousis, R.~Gregory and V.~A.~Rubakov,
%``Wave function of the radion in a brane world,''
Phys.\ Rev.\ D {\bf 62}, 067505 (2000) [hep-th/9912160].

\end{thebibliography}
\end{document}